\documentclass[12pt]{article}
\usepackage{amssymb}
\usepackage{amsfonts}
\usepackage{epsfig}
\usepackage{graphicx}
\usepackage{epsfig}
\usepackage{xcolor}

\begin{document}

\title{The Topology of African Exports: emerging patterns on spanning trees}
\author{Tanya Ara\'{u}jo and M. Ennes Ferreira \and ISEG, University of
Lisbon \and {\small Miguel Lupi 20, 1248-079 Lisbon} \and {\small UECE -
Research Unit on Complexity and Economics}}
\date{}
\maketitle

\begin{abstract}
This paper is a contribution to interweaving two lines of research that have
progressed in separate ways: network analyses of international trade and the
literature on African trade and development. Gathering empirical data on
African countries has important limitations and so does the space occupied
by African countries in the analyses of trade networks. Here, these
limitations are dealt with by the definition of two independent bipartite
networks: a destination share network and\ a\ commodity share network.

These networks - together with their corresponding minimal spanning trees -
allow to uncover some ordering emerging from African exports in the broader
context of international trade. The emerging patterns help to understand
important characteristics of African exports and its binding relations to
other economic, geographic and organizational concerns as the recent
literature on African trade, development and growth has shown.
\end{abstract}

\bigskip *Financial support from national funds by FCT (Funda\c{c}\~{a}o
para a Ci\^{e}ncia e a Tecnologia). This article is part of the Strategic
Project: UID/ECO/00436/2013

\bigskip

Keywords: Trade networks, African exports, Spanning trees, Bipartite graphs

\bigskip

\section{Introduction}

A growing literature has presented empirical findings of the persistent
impact of trade activities on economic growth and poverty reduction (\cite%
{Portugal-Perez},\cite{Kamuganga},\cite{Morrissey},\cite{Baliamoune},\cite%
{Ackah}, \cite{Gamberoni}). Besides discussing on the relation between trade
and development, they also report on the growth by destination hypothesis,
according to which, the destination of exports can play an important role in
determining the trade pattern of a country and its development path.

Simultaneously, there has been a growing interest in applying concepts and
tools of network theory to the analysis of international trade (\cite%
{Serrano},\cite{Almog},\cite{Fagiolo},\cite{Saracco}, \cite{Benedictis},\cite%
{Picciolo},\cite{Yang}). Trade networks are among the most cited examples of
the use of network approaches. The international trade activity is an
appealing example of a large-scale system whose underlying structure can be
represented by a set of bilateral relations.

This paper is a contribution to interweaving two lines of research that have
progressed in separate ways: network analyses of international trade and the
literature on African trade and development.

The most intuitive way of defining a trade network is representing each
world country by a vertex and the flow of imports/exports between them by a
directed link. Such descriptions of bilateral trade relations have been used
in the gravity models (\cite{Bergstrand}) where some structural and
dynamical aspects of trade have been often accounted for.

While some authors have used network approaches to investigate the
international trade activity, studies that apply network models to focus on
specific issues of African trade are less prominent. Although African
countries are usually considered in international trade network analyses,
the space they occupy in these literature is often very narrow.

This must be partly due to the existence of some relevant limitations that
empirical data on African countries suffer from, mostly because part of
African countries does not report trade data to the United Nations. The
usual solution in this case is to use partner country data, an approach
referred to as \textbf{mirror statistics}. However, using mirror statistics
is not a suitable source for bilateral trade in Africa as an important part
of intra-African trade concerns import and exports by non-reporting
countries.

A possible solution to overcome the limitations on bilateral trade data is
to make use of information that, although concerning two specific trading
countries, might be provided indirectly by a third and secondary source.
That is what happens when we define a bipartite network and its one-mode
projection. In so doing, each bilateral relation between two African
countries in the network is defined from the relations each of these
countries hold with another entity. It can be achieved in such a way that
when they are similar enough in their relation with that other entity, a
link is defined between them.

Our approach is applied to a subset of 49 African countries and based on the
definition of two independent bipartite networks where trade similarities
between each pair of African countries are used to define the existence of a
link. In the first bipartite graph, the similarities concern a mutual
leading destination of exports by each pair of countries and in the second
bipartite graph, countries are linked through the existence of a mutual
leading export commodity between them.

Therefore, bilateral trade discrepancies are avoided and we are able to look
simultaneously at network structures that emerge from two fundamental
characteristics (exporting destinations and exporting commodities) of\ the
international trade. As both networks were defined from empirical data
reported for 2014, we call these networks \textbf{%
destination share networks} \textbf{\ } (DSN$_{14}$) and\textbf{%
\ } \textbf{commodity share networks} (CSN$%
_{14})$, respectively.

Its worth noticing that the choice of a given network representation is only
one out of several other ways to look at a given system. There may be many
ways in which the elementary units and the links between them are conceived
and the choices may depend strongly on the available empirical data and on
the questions that a network analysis aims to address (\cite{Araujo2014}).

The main question addressed in this paper is whether some relevant
characteristics of African trade would emerge from the bipartite networks
above described. We hypothesized that specific characteristics could come
out and shape the structures of both the DSN$_{14}$ and the CSN$_{14}$. We
envision that these networks will allow to uncover some ordering emerging
from African exports in the broader context of international trade. If it
happens, the emerging patterns may help to understand important
characteristics of African exports and its relation to other economic,
geographic and organizational concerns.

To this end, the paper is organized as follows: next section presents the
empirical data we work with, Section three describes the methodology and
some preliminary results from its application. In Section four we present
further results and discuss on their interpretation in the international
trade setting. Section five concludes and outlines future work.\newpage

\section{Data}

Trade Map - Trade statistics for international business development (\cite%
{ITM}) - provides a dataset of import and export data in the form of tables,
graphs and maps for a set of reporting and non-reporting countries all over
the world. There are also indicators on export performance, international
demand, alternative markets and competitive markets. Trade Map covers 220
countries and territories and 5300 products of the Harmonized System (HS
code).

Since the Trade Map statistics capture nationally reported data of such a
large amount of countries, this dataset is an appropriate source to the
empirical study of temporal patterns emerging from international trade.
Nevertheless, some major limitations should be indicated, as for countries
that do not report trade data to the United Nations, Trade Map uses partner
country data, an issue that motivated our choice for defining bipartite
networks.

Our approach is applied to a subset of 49 African countries (see Table 1)
and from this data source, trade similarities between each pair of countries
are used to define networks of links between countries.

Table 1 shows the 49 African countries we have been working with. It also
shows the regional organization of each country, accordingly to the
following classification: 1 - Southern African Development Community (SADC);
2 - Uni\~{a}o do Magreb \'{A}rabe (UMA); 3 - Comunidade Econ\'{o}mica dos
Estados da Africa Central (CEEAC); 4 - Common Market for Eastern and
Southern Africa (COMESA) and 5 - Comunidade Econ\'{o}mica dos Estados da
\'{A}frica Ocidental (CEDEAO).

For each African country in Table 1, we consider the set of countries to
which at least one of the African countries had exported in the year of
2014. The specification of the destinations of exports of each country
followed the International Trade Statistics database (\cite{ITM}) from where
just \textbf{the first and the second main destinations of exports} \textbf{%
of each country} were taken.

Similarly and also for each African country in Table 1, we took the set of
commodities that at least one of the African countries had exported in 2014.
The specification of the destinations of exports of each country followed
the same database from where just \textbf{the first and the second main
export commodities of each country} were taken.

For each country (column label "\textbf{Country}") in Table 1, besides the regional organization (column label
"\textbf{O}") and the first and second
destinations (column labels "\textbf{Destinations}") and commodities (column labels "\textbf{%
Products}"), we also considered the export value in 2014 (as
reported in \cite{ITM}) so that the size of the representation of each
country in the networks herein presented is proportional to its
corresponding export value in 2014.

\begin{center}
\begin{tabular}{|l|l|l|l|l|l|l|l|l|l|l|l|l|l|}
\hline
\multicolumn{2}{|c|}{\tiny Country} & {\tiny O} & \multicolumn{2}{|l}{\tiny %
Destinations} & \multicolumn{2}{|l}{\tiny Products} & \multicolumn{2}{|c|}%
{\tiny Country} & {\tiny O} & \multicolumn{2}{|l|}{\tiny Destinations} &
\multicolumn{2}{|l|}{\tiny Products} \\ \hline
{\tiny Seychelles} & {\tiny SEY} & {\tiny 1} & {\tiny FRA} & {\tiny UK} &
{\tiny 16} & {\tiny 3} & {\tiny Gabon} & {\tiny GAB} & {\tiny 3} & {\tiny CHI%
} & {\tiny JAP} & {\tiny 27} & {\tiny 44} \\ \hline
{\tiny Angola} & {\tiny AGO} & {\tiny 1} & {\tiny CHI} & {\tiny USA} &
{\tiny 27} & {\tiny 71} & {\tiny Burundi} & {\tiny BUR} & {\tiny 3} & {\tiny %
PAK} & {\tiny RWA} & {\tiny 9} & {\tiny 71} \\ \hline
{\tiny Mozambique} & {\tiny MOZ} & {\tiny 1} & {\tiny CHI} & {\tiny ZAF} &
{\tiny 27} & {\tiny 76} & {\tiny St.Tome} & {\tiny STP} & {\tiny 3} & {\tiny %
BEL} & {\tiny TUR} & {\tiny 18} & {\tiny 10} \\ \hline
{\tiny D.R.Congo} & {\tiny DRC} & {\tiny 1} & {\tiny CHI} & {\tiny ZMB} &
{\tiny 74} & {\tiny 26} & {\tiny Kenya} & {\tiny KEN} & {\tiny 4} & {\tiny %
ZMB} & {\tiny TZA} & {\tiny 9} & {\tiny 6} \\ \hline
{\tiny Botswana} & {\tiny BWT} & {\tiny 1} & {\tiny BEL} & {\tiny IND} &
{\tiny 71} & {\tiny 75} & {\tiny Egypto} & {\tiny EGY} & {\tiny 4} & {\tiny %
ITA} & {\tiny GER} & {\tiny 27} & {\tiny 85} \\ \hline
{\tiny South Africa} & {\tiny ZAF} & {\tiny 1} & {\tiny CHI} & {\tiny USA} &
{\tiny 71} & {\tiny 26} & {\tiny Ethiopia} & {\tiny ETH} & {\tiny 4} &
{\tiny CHI} & {\tiny SWI} & {\tiny 27} & {\tiny 9} \\ \hline
{\tiny Zambia} & {\tiny ZAM} & {\tiny 1} & {\tiny CHI} & {\tiny KOR} &
{\tiny 74} & {\tiny 28} & {\tiny Uganda} & {\tiny UGA} & {\tiny 4} & {\tiny %
RWA} & {\tiny NET} & {\tiny 9} & {\tiny 27} \\ \hline
{\tiny Tanzania} & {\tiny TZA} & {\tiny 1} & {\tiny IND} & {\tiny CHI} &
{\tiny 71} & {\tiny 26} & {\tiny Eritrea} & {\tiny ERI} & {\tiny 4} & {\tiny %
CHI} & {\tiny IND} & {\tiny 26} & {\tiny 9} \\ \hline
{\tiny Namibia} & {\tiny NAM} & {\tiny 1} & {\tiny BWA} & {\tiny ZAF} &
{\tiny 71} & {\tiny 3} & {\tiny Comoros} & {\tiny NGA} & {\tiny 4} & {\tiny %
IND} & {\tiny GER} & {\tiny 9} & {\tiny 89} \\ \hline
{\tiny Zimbabwe} & {\tiny ZWE} & {\tiny 1} & {\tiny CHI} & {\tiny ZAF} &
{\tiny 71} & {\tiny 24} & {\tiny Rwanda} & {\tiny RWA} & {\tiny 4} & {\tiny %
CHI} & {\tiny MAS} & {\tiny 26} & {\tiny 9} \\ \hline
{\tiny Mauritius} & {\tiny MUS} & {\tiny 1} & {\tiny USA} & {\tiny FRA} &
{\tiny 61} & {\tiny 62} & {\tiny Guine Bissau} & {\tiny GuB} & {\tiny 5} &
{\tiny IND} & {\tiny CHI} & {\tiny 8} & {\tiny 44} \\ \hline
{\tiny Lesotho} & {\tiny LES} & {\tiny 1} & {\tiny USA} & {\tiny BEL} &
{\tiny 71} & {\tiny 61} & {\tiny Ghana} & {\tiny GHA} & {\tiny 5} & {\tiny %
ZAF} & {\tiny EMI} & {\tiny 27} & {\tiny 18} \\ \hline
{\tiny Malawi} & {\tiny MWI} & {\tiny 1} & {\tiny BEL} & {\tiny GER} &
{\tiny 24} & {\tiny 12} & {\tiny Cote d'Ivoire} & {\tiny CIV} & {\tiny 5} &
{\tiny USA} & {\tiny GER} & {\tiny 18} & {\tiny 27} \\ \hline
{\tiny Swaziland} & {\tiny SWA} & {\tiny 1} & {\tiny ZAF} & {\tiny IND} &
{\tiny 33} & {\tiny 17} & {\tiny Nigeria} & {\tiny NGA} & {\tiny 5} & {\tiny %
IND} & {\tiny BRA} & {\tiny 27} & {\tiny 18} \\ \hline
{\tiny Madagascar} & {\tiny MDG} & {\tiny 1} & {\tiny FRA} & {\tiny USA} &
{\tiny 75} & {\tiny 9} & {\tiny Burkina Faso} & {\tiny BFA} & {\tiny 5} &
{\tiny SWI} & {\tiny CHI} & {\tiny 71} & {\tiny 52} \\ \hline
{\tiny Algeria} & {\tiny MDG} & {\tiny 2} & {\tiny ESP} & {\tiny ITA} &
{\tiny 27} & {\tiny 28} & {\tiny Senegal} & {\tiny SEN} & {\tiny 5} & {\tiny %
SWI} & {\tiny IND} & {\tiny 27} & {\tiny 3} \\ \hline
{\tiny Lybia} & {\tiny LYB} & {\tiny 2} & {\tiny ITA} & {\tiny FRA} & {\tiny %
27} & {\tiny 72} & {\tiny Benin} & {\tiny BEN} & {\tiny 5} & {\tiny BFA} &
{\tiny CHI} & {\tiny 52} & {\tiny 27} \\ \hline
{\tiny Morocco} & {\tiny MAR} & {\tiny 2} & {\tiny ESP} & {\tiny FRA} &
{\tiny 85} & {\tiny 87} & {\tiny Liberia} & {\tiny LIB} & {\tiny 5} & {\tiny %
CHI} & {\tiny POL} & {\tiny 26} & {\tiny 89} \\ \hline
{\tiny Tunisia} & {\tiny TUN} & {\tiny 2} & {\tiny FRA} & {\tiny ITA} &
{\tiny 85} & {\tiny 62} & {\tiny Mali} & {\tiny MAL} & {\tiny 5} & {\tiny SWI%
} & {\tiny CHI} & {\tiny 52} & {\tiny 71} \\ \hline
{\tiny Mauritania} & {\tiny MRT} & {\tiny 2} & {\tiny CHI} & {\tiny SWI} &
{\tiny 26} & {\tiny 3} & {\tiny Niger} & {\tiny NIG} & {\tiny 5} & {\tiny FRA%
} & {\tiny BFA} & {\tiny 26} & {\tiny 27} \\ \hline
{\tiny Cameroon} & {\tiny CMR} & {\tiny 3} & {\tiny ESP} & {\tiny CHI} &
{\tiny 27} & {\tiny 18} & {\tiny Togo} & {\tiny TOG} & {\tiny 5} & {\tiny BFA%
} & {\tiny LEB} & {\tiny 52} & {\tiny 39} \\ \hline
{\tiny Chad} & {\tiny CHA} & {\tiny 3} & {\tiny USA} & {\tiny JAP} & {\tiny %
27} & {\tiny 52} & {\tiny Sierra Leone} & {\tiny SLe} & {\tiny 5} & {\tiny %
CHI} & {\tiny BEL} & {\tiny 26} & {\tiny 71} \\ \hline
{\tiny C.African R.} & {\tiny CAR} & {\tiny 3} & {\tiny CHI} & {\tiny IDN} &
{\tiny 44} & {\tiny 52} & {\tiny Cabo Verde} & {\tiny CaV} & {\tiny 5} &
{\tiny ESP} & {\tiny POR} & {\tiny 3} & {\tiny 16} \\ \hline
{\tiny Congo} & {\tiny COG} & {\tiny 3} & {\tiny CHI} & {\tiny ITA} & {\tiny %
27} & {\tiny 89} & {\tiny Guinea} & {\tiny GIN} & {\tiny 5} & {\tiny KOR} &
{\tiny IND} & {\tiny 27} & {\tiny 26} \\ \hline
{\tiny Eq.Guine} & {\tiny EqG} & {\tiny 3} & {\tiny CHI} & {\tiny UK} &
{\tiny 27} & {\tiny 29} &  &  &  &  &  &  &  \\ \hline
\end{tabular}

{\small Table 1: African countries and their classification into regional
organizations, their main exporting commodities and their leading
destinations of exports in 2014. Source: International Trade Map
(http://www.trademap.org) (\cite{ITM}).}
\end{center}

\subsection{The Destinations of Exports}

The following list of 28 countries (Countries$_{14}$)\ that imported from
Africa in 2014 on a first and second destination basis (as just the first
and the second main destinations of exports of each country were taken) are
grouped in five partition clusters: "\textbf{African Countries" , "USA", "China" ,
"Europe"} and \textbf{\ "Other"}.

\begin{enumerate}
\item African Countries

\begin{tabular}{|l|l|l|l|l|l|}
\hline
{\tiny Zambia(ZMB)} & {\tiny Tanzania(TZA)} & {\tiny Botswana(BWA)} & {\tiny %
\ South Africa(ZAF)} & {\tiny Rwanda(RWA)} & {\tiny B.Faso(BFA)} \\ \hline
\end{tabular}

\item USA

\item China

\item Europe

\begin{tabular}{|l|l|l|l|l|}
\hline
{\tiny France(FRA)} & {\tiny Switzerland(SWI)} & {\tiny Netherlands(NET)} &
{\tiny Italy(ITA)} & {\tiny Poland(POL)} \\ \hline
{\tiny United Kindom(UK)} & {\tiny Spain(ESP)} & {\tiny Portugal(POR)} &
{\tiny Belgium(BEL)} & {\tiny Germany(GER)} \\ \hline
\end{tabular}

\item Other

\begin{tabular}{|l|l|l|l|l|}
\hline
{\tiny Malaysia(MAS)} & {\tiny India(IND)} & {\tiny Emirates(EMI)} & {\tiny %
Turkey(TUR)} & {\tiny Brazil(BRA)} \\ \hline
{\tiny Korea(KOR)} & {\tiny Japan(JAP)} & {\tiny Indonesia(IDN)} & {\tiny %
Lebanon(LEB)} & {\tiny Pakistan(PAK)} \\ \hline
\end{tabular}
\end{enumerate}

\medskip

Figure 1 shows the distribution of the frequencies of the two leading
destinations of exports of each country in Table 1. The first histogram
(right) in Figure 1 allows for the observation of the leading destinations
of exports from Africa in 2014 and to the way they are distributed by
countries. It also shows the distribution of the frequencies (left plot) of
the first and second destinations when frequencies are aggregated in the
five partition clusters above described.

The first plot shows that the top-five destinations of African exports in
2014 were China, South Africa, Switzerland, France and India. China holds
the highest frequency, being followed by India and by two EU countries
(Switzerland and France). The second histogram shows that when frequencies
are aggregated in five partition clusters, "Europe" holds the highest frequency, being followed by
"China".

\begin{center}
\begin{figure}[htb]
\psfig{figure=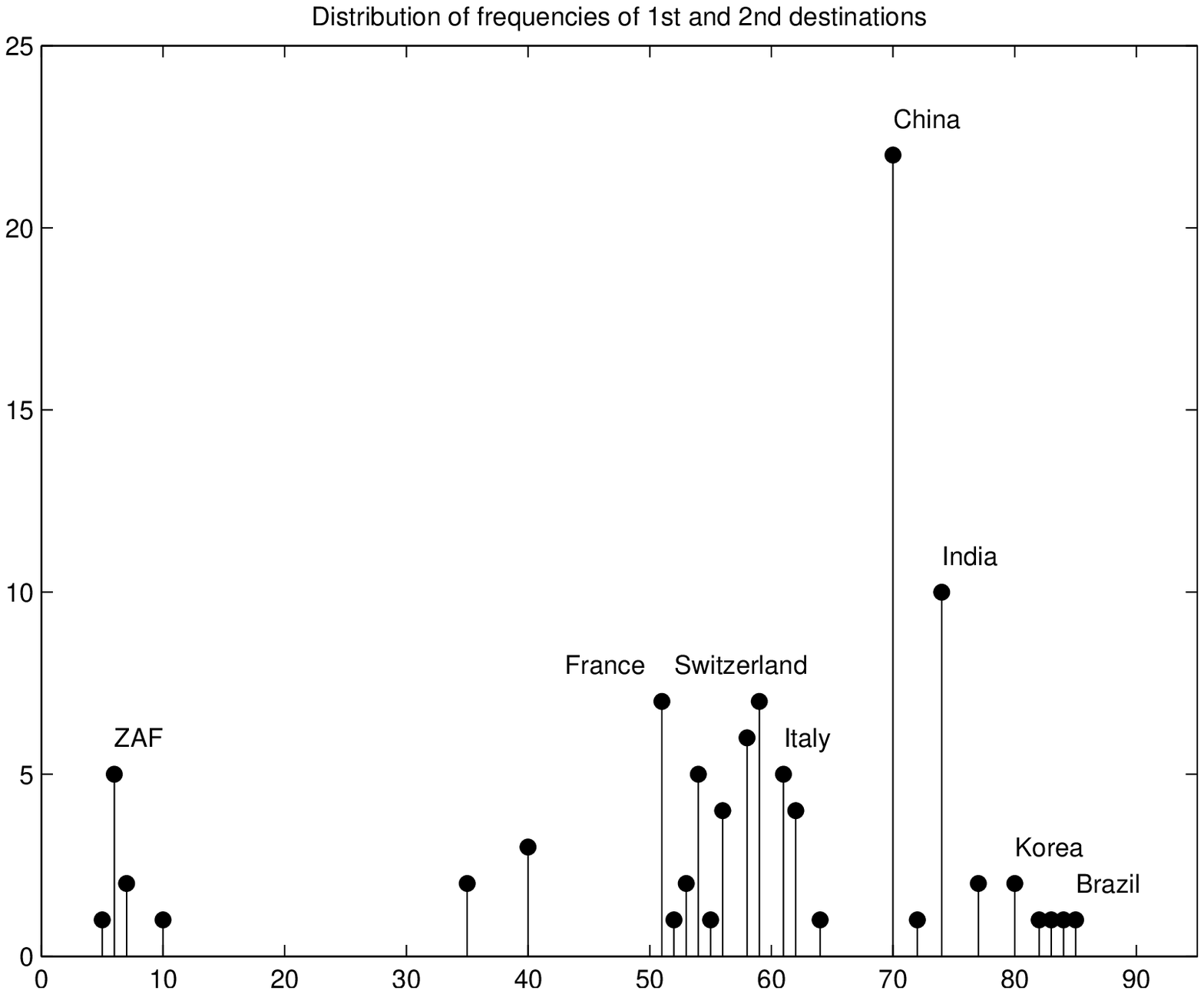,width=6.5truecm}
\psfig{figure=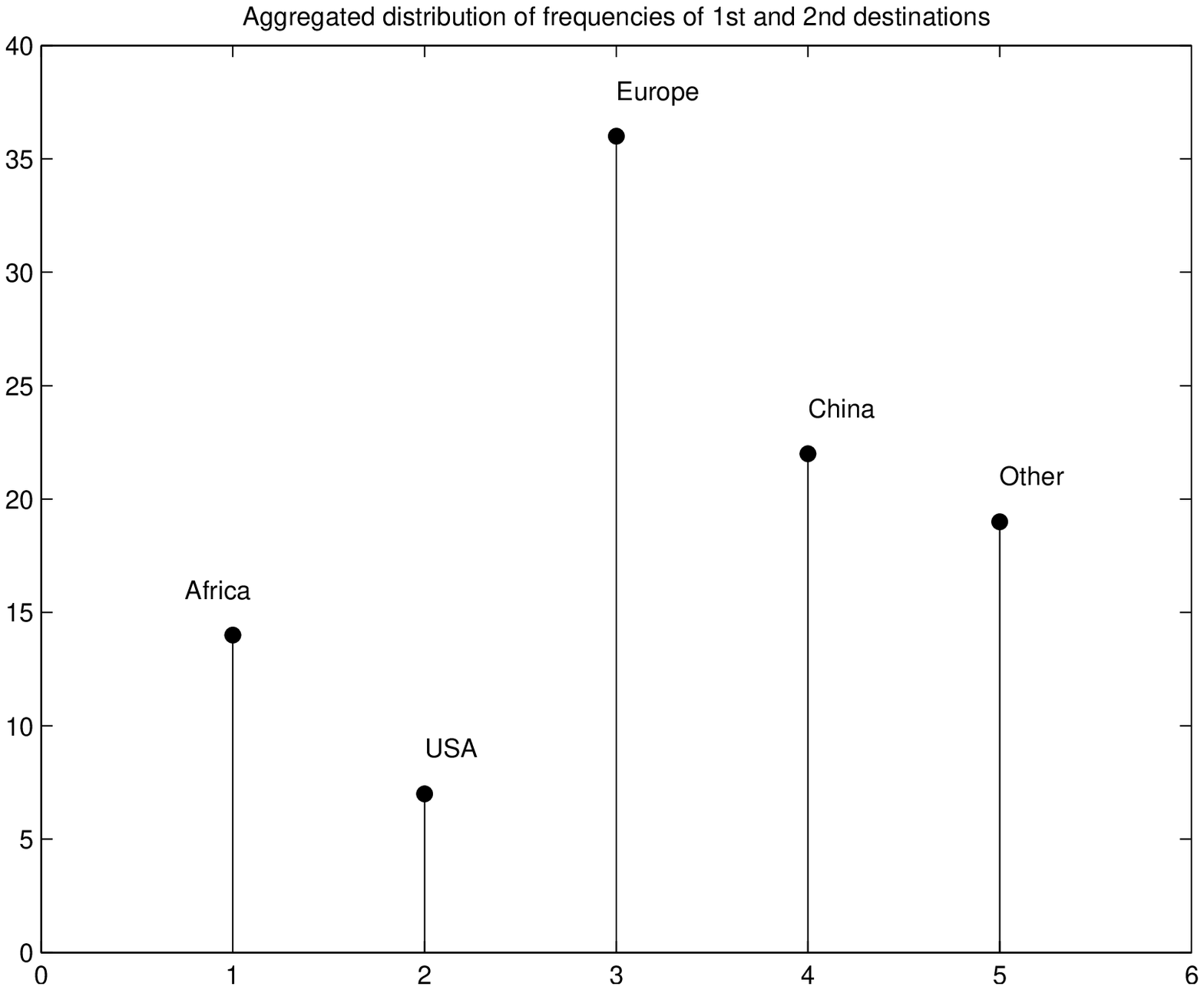,width=6.5truecm}
\caption{The distribution of the frequencies of the two leading
destinations by country and the same distribution when destinations are
aggregated in five partition clusters.}
\end{figure}
\end{center}

\subsection{The Exporting Commodities}

The following list of 27 commodities (Commodities$_{14}$) imported from
Africa in 2014 on a first and second product basis (as just the first and
the second main export commodities of each country were taken) are
aggregated in five partition clusters: "\textbf{Petroleum", "Raw Materials",
"Diamonds", "Manufactured Products"} and \textbf{"Other Raw Materials"}.

\begin{enumerate}
\item Petroleum: HS code:\ 27(Oil Fuels)

\item Raw Materials (HS code)

\begin{tabular}{|c|c|c|c|c|c|}
\hline
{\tiny 03 (fish)} & {\tiny 06(trees)} & {\tiny 08(fruit)} & {\tiny 09(coffee)%
} & {\tiny 10(cereals)} & {\tiny 16(meat)} \\ \hline
{\tiny 17(sugars)} & {\tiny 18(cocoa)} & {\tiny 24(tobacco)} & {\tiny %
33(oils)} & {\tiny 44(wood)} & {\tiny 52(cotton)} \\ \hline
\end{tabular}

\item Diamonds: HS code:\ 71(Pearls)

\item Manufactured Products (HS code)

\begin{tabular}{|l|l|l|l|l|l|}
\hline
{\tiny 28(inorg.chemic.)} & {\tiny 29(org.chemic.)} & {\tiny 39(plastics)} &
{\tiny 61(art.apparel)} & {\tiny 62(art.apparel)} & {\tiny 72(iron-steel)}
\\ \hline
{\tiny 74(copper)} & {\tiny 75(nickel)} & {\tiny 76(Aluminium)} & {\tiny %
85(electricals)} & {\tiny 87(vehicles)} & {\tiny 89(boats)} \\ \hline
\end{tabular}

\item Other Raw Materials: HS code:\ 26(Ores)
\end{enumerate}

Figure 2 shows the distribution of the frequencies of the two leading export
products of each country in Table 1. The second plot (left) shows the same
distribution when the products are aggregated according to the five
partitions of commodities above presented.

\begin{center}
\begin{figure}[htb]
\psfig{figure=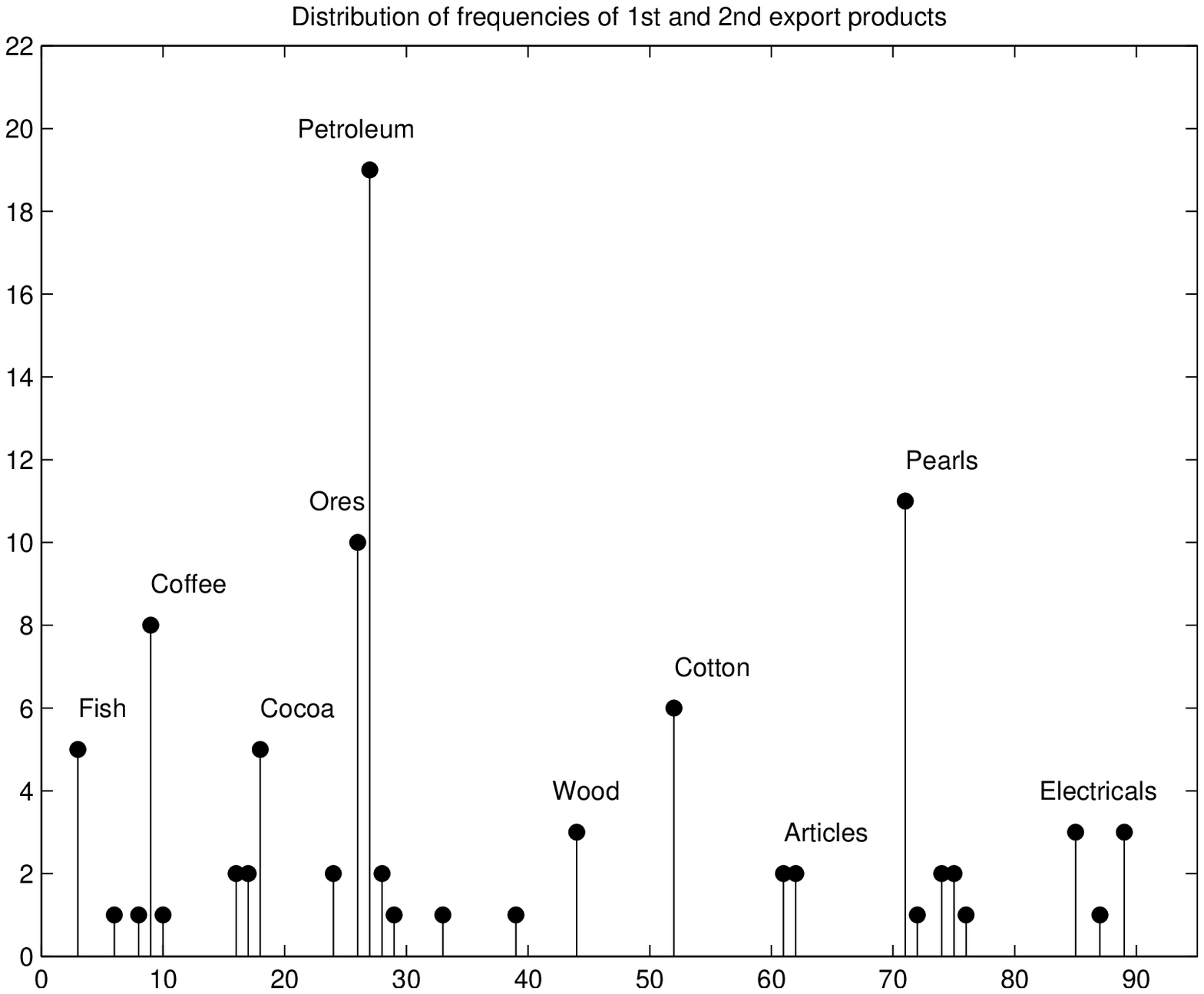,width=6.5truecm}
\psfig{figure=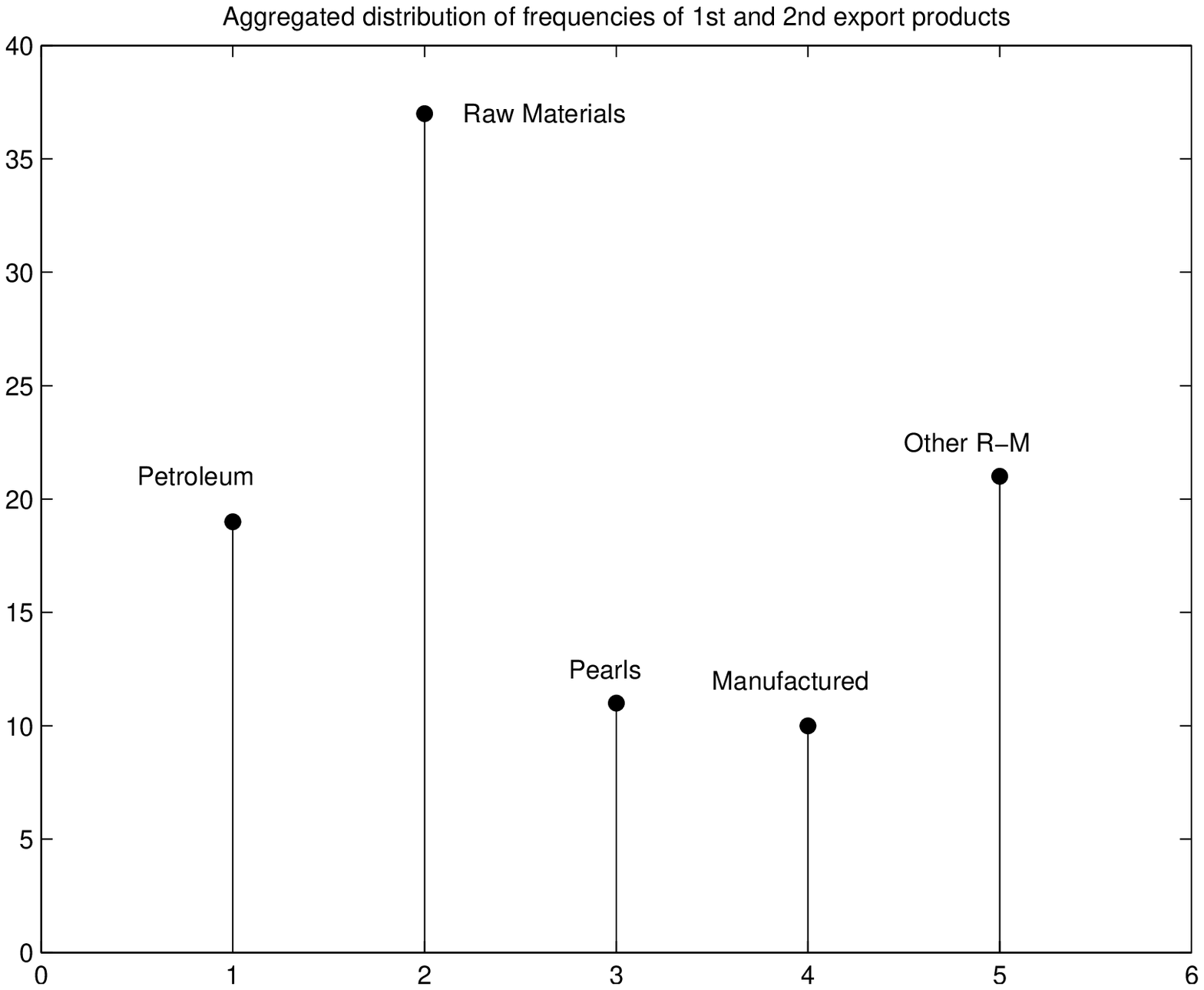,width=6.5truecm}
\caption{The distribution of the frequencies of the two leading
export commodities by country and the same distribution when commodities are
aggregated in five partition clusters.}
\end{figure}
\end{center}

From the first histogram in Figure 2, one is able to observe the most
requested products exported from Africa in 2014 and to the way their
frequencies are distributed. Not surprisingly, Petroleum crude holds the
highest frequency, being followed by Pearls, Ores and by Coffee. The
top-five exporting commodities in 2014 when frequencies are aggregated are
"Raw Materials", and "Other Raw
Materials", with which "Petroleum" shares a similar frequency.

\section{\protect\bigskip Methodology}

Network-based approaches are nowadays quite common in the analysis of
systems where a network representation intuitively emerges. It often happens
in the study of international trade networks.

As earlier mentioned, the choice of a given network representation is only
one out of several other ways to look at a given system. There may be many
ways in which the elementary units and the links between them are defined.
Here we define two independent bipartite networks where trade similarities
between each pair of countries are used to define the existence of every
link in each of those networks.

In this section, we analyze the projections of those bipartite networks, the
earlier described DSN$_{14}$ and CSN$_{14}$ networks are weighted graphs and
the weight of each link corresponds to the intensity of the similarity
between the linked pair of countries. In the next section, the weighted
networks are further analyzed through the construction of their
corresponding minimal spanning tress (MST). In so doing, we are able to
emphasize the main topological patterns that emerge from the network
representations and to discuss their interpretation in the international
trade setting.

\subsection{Bipartite Graphs}

A bipartite network $B$ consists of two partitions of nodes $V$ and $W$,
such that edges connect nodes from different partitions, but never those in
the same partition. A one-mode projection of such a bipartite network onto $%
V $ is a network consisting of the nodes in $V$; two nodes $v$ and $v\prime $
are connected in the one-mode projection, if and only if there exist a node $%
w\in W$ such that $(v,w)$ and $(v\prime ,w)$ are edges in the corresponding
bipartite network ($B$).

In the following, we explore two bipartite networks and their corresponding
one-mode projections, the earlier described DSN$_{14}$ and CSN$_{14}$
networks.

\subsubsection{Topological Coefficients}

The adoption of a network approach provides well-known notions of graph
theory to fully characterize the structure of the projections DSN$_{14}$ and
CSN$_{14}$. These notions are formally defined as topological coefficients.
Here, we concentrate on the calculation of five coefficients. Three of them
are quantities related to averages values of one topological coefficient
defined at the node level, as the network degree $\left\langle
k\right\rangle $, the betweenness centrality $\left\langle B\right\rangle $
and the average clustering coefficient $\left\langle C\right\rangle $. The
other two coefficients are measured at the network level, they are the
density ($d$) of the network and the network diameter ($D$).

\begin{enumerate}
\item the average degree ($\left\langle k\right\rangle $) of a network
measures the average number of links connecting each element of the network.

\item the betweenness centrality ($\left\langle B\right\rangle $) measured
as the fraction of paths connecting all pairs of nodes and containing the
node of interest ($i$).

\item the clustering coefficient ($\left\langle C\right\rangle $) measures
the average probability that two nodes having a common neighbor are
themselves connected%
\begin{equation}
C_{i}=\frac{E(v_{i})}{v_{i}(v_{i}-1)}  \label{1}
\end{equation}
\end{enumerate}

where $E(v_{i})$ is the size of the neighbourhood ($v_{i}$) of the node $i$
and the neighbourhood of $i$ consisting of all nodes adjacent to $i$.

\begin{enumerate}
\item[4.] the diameter of the network ($\left\langle D\right\rangle $)
measuring the shortest distance between the two most distant nodes in the
network.

\item[5.] the density ($0\leq d\leq 1$) of the network is the ratio of the
number of links in the network to the number of possible links%
\begin{equation}
d=\frac{2L}{n(n-1)}  \label{2}
\end{equation}
\end{enumerate}

where $L$ is the number of links and $n$ is the number of nodes.

Here, these coefficients are computed for different sub-graphs of both the
DSN$_{14}$ and the CSN$_{14}$. The nodes (countries) in each sub-graph are
grouped accordingly to the partition clusters of main destinations (%
"African Countries", "USA",
"China", "Europe"
and "Other") and by partition of main exporting
commodity ("Petroleum", "Raw
Materials", "Diamonds",
"Manufactured Products" and "Other Raw Materials"). We also apply these measures to the
partition clusters defined by the regional organizations to which the
countries belong (SADC, UMA, CEEAC, COMESA and CEDEAO). In so doing, it is
possible to compare in terms of topological coefficients the different
structures of the \ DSN$_{14}$ and the CSN$_{14}$ networks. For each of
them, the topological coefficients are computed at the node level and then
averaged by partitions of interest (main destination, main commodity or
regional organization).

\subsection{First Results}

As a first approach and from the 49 African countries in Table 1 we start by
developing the DSN$_{14}$ of 2014. Then, and from the same set of countries
and the same year, we develop the CSN$_{14}$.

\subsubsection{Connecting countries by a mutual export destination\allowbreak%
}

The bipartite network DSN$_{14}$ consists of the following partitions:

\begin{itemize}
\item the set $A$ of 49 African countries presented in Table 1 and

\item the set of Countries$_{14}$ (Section 2.1) to which at least one of the
countries in $A$ had exported in 2014 on a first and second main destination
basis.
\end{itemize}

As such, in the DSN$_{14}$ two countries are linked if and only if they
shared a mutual destination of exports in 2014 among their two main export
destinations. We have considered the two main destinations of exports of
each country in Table 1 (columns "Destinations"). Otherwise, if just the main destination was considered, the resulting DSN$%
_{14}$ would comprise a set of disconnected and complete sub-graphs as, by
definition, each country has just one main destination. Therefore, links in
the DSN$_{14}$ are weighted by the number of coincident destinations a pair
of countries share (among the two main destinations of each country in the
pair), consequently, every link $L_{(i,j)}$ in the DSN$_{14}$ takes value in
the set $\left\{ 0,1,2\right\} $.

As an example, $L_{(AGO,ZAF)}=2$ since AGO and ZAF share two main
destinations of exports in 2014: China and USA.

Another example is $L_{(CMR,CAV)}=1$ due to CMR and CaV mutual destination
of exports to ESP in 2014.

Among the many examples of missing links there are the cases of AGO and KEN (%
$L_{(AGO,KEN)}=0$) and\ AGO and CaV ($L_{(AGO,CAV)}=0$) since neither AGO
and KEN nor AGO and CaV share any mutual leading destination of exports in
2014.

Figure 3 presents the DSN$_{14}$, a network of 49 African countries linked
by mutual leading destinations of exports in 2014. Nodes are colored
according to the partition cluster to which their main export destination
belongs: red nodes identify countries whose main export destination is
"China", yellow for "Europe", green for "USA", blue for
"African countries" and purple for the cluster
of "Other".

Such a partition of the set of countries into five clusters - defined by the
country main destination of exports - allows for computing the average
values of some topological coefficients by partition cluster.

In so doing, and as we shall see later in the paper, it is possible to
compare important patterns coming out from the DSN$_{14}$ and to evaluate
the extent to which the emerging patterns relate to geographic, regional or
economic concerns.

\begin{center}
\begin{figure}[htb]
\psfig{figure=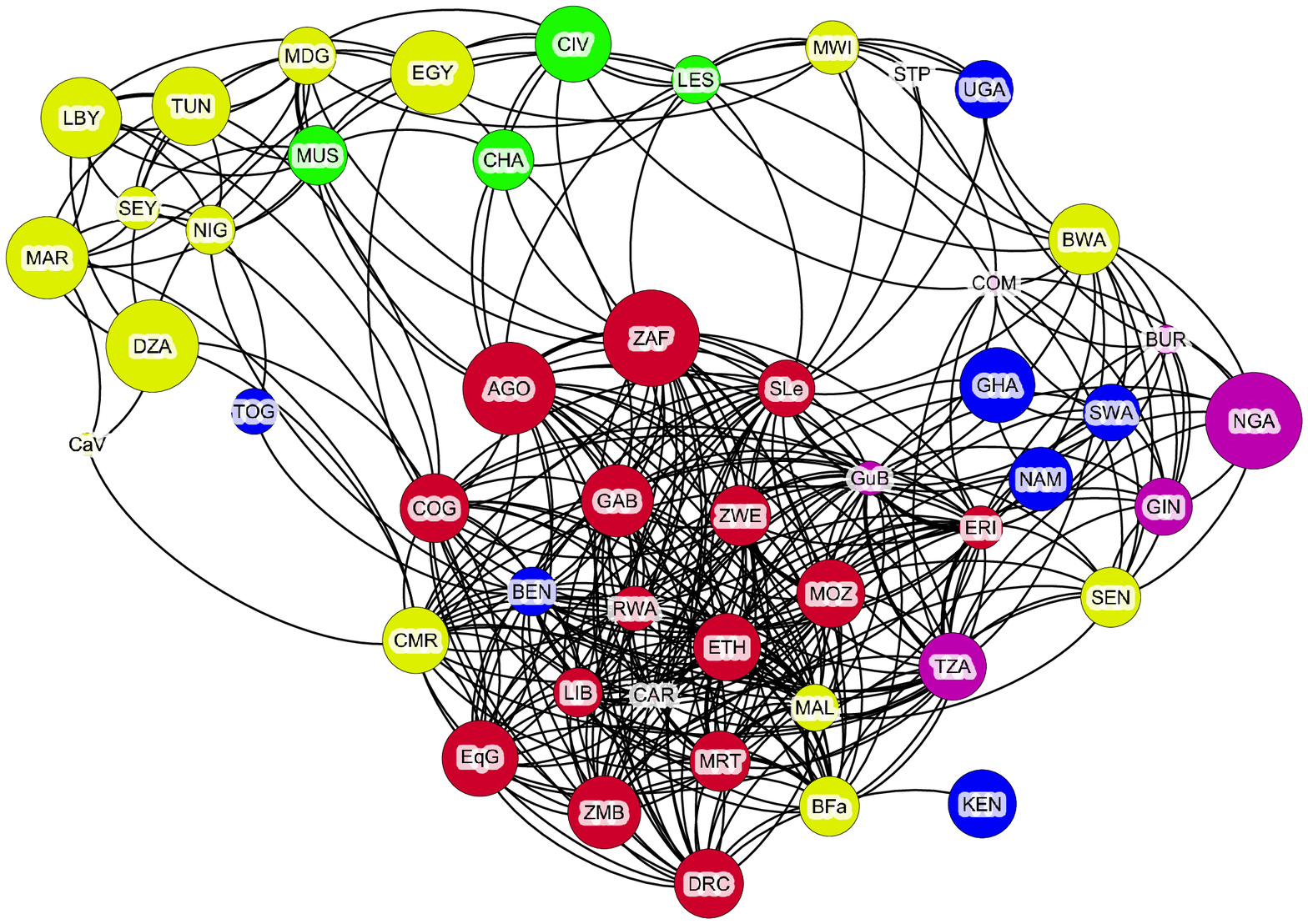,width=12truecm}
\caption{The DSN$_{14}$ colored by partition of main destination.}
\end{figure}
\end{center}

In all networks presented in this paper, the size of each node is
proportional to the export value of the country in 2014. Therefore, the
largest nodes are ZAF, AGO, DZA and NGA since these countries hold the
largest amounts of export values in 2014. Figure 3 shows that the highest
connected nodes are those whose main export destination is "%
China"(colored red), showing that these countries are those
that share (with other countries in the whole network) the highest numbers
of mutual destinations of exports. On the other hand, there are countries
like TOG and STP that share very few mutual destinations with any other
countries in the network.

The remarkable (red) bulk of highly connected countries in the
"China" destination cluster is the very first
pattern coming out from our approach. It is followed by another interesting
result that clusters together the exporters to "Europe%
"(colored yellow), in the left upper side of Figure 3. There,
another (small) cluster of exporters to "USA"%
(colored green) can also be seen. On the other hand, countries that have
"African countries"(purple) as the first
destination of exports seem to be the less clustered in the DSN$_{14}$,
being followed by those that export mainly to countries in the partition of
"Other"(colored blue) like TOG and STP.

Another evidence coming out from the network in Figure 3 is that, excluding
NGA, the strongest connected countries coincide with the countries with the
highest amounts of export values in 2014 (the larger nodes). Not
surprisingly, it shows a positive non-negligible correlation between the
amount of exports of a country and its weighted degree in the DSN$_{14}$:
the countries with the highest amounts of exports in 2014 tend to be those
that cluster as exporters to the most frequent African export destinations:
"China" and "Europe%
"(most of the large nodes are yellow and red nodes).

Table 2 shows some topological coefficients computed for each node of the DSN%
$_{14}$ and averaged by partition of main destination. The second column (%
"\textbf{Size}" ) shows the number of countries
in each partition. The averages of the weighted degree ($\left\langle
k\right\rangle $), betweenness ($\left\langle B\right\rangle $) and
clustering ($\left\langle C\right\rangle $) of each partition show that
there is a remarkable clustering in the "China"%
partition and that although the "Europe"%
cluster has the same number of nodes, its values of $\left\langle
k\right\rangle ,\left\langle B\right\rangle $ and mostly $\left\langle
C\right\rangle $ are far below those of the "China%
"partition, confirming the relevance of the partition of
"China"exporters as the very first pattern
coming out from our approach.

\begin{center}
\begin{tabular}{|l|l|l|l|l|}
\hline
{\small Partition} & {\small Size} & $\left\langle k\right\rangle $ & $%
\left\langle B\right\rangle $ & $\left\langle C\right\rangle $ \\ \hline
\multicolumn{1}{|r|}{${\small Africa}$} & \multicolumn{1}{|r|}{${\small 7}$}
& \multicolumn{1}{|r|}{${\small 15.8}$} & \multicolumn{1}{|r|}{${\small 22.3}
$} & \multicolumn{1}{|r|}{${\small 0.78}$} \\ \hline
\multicolumn{1}{|r|}{${\small USA}$} & \multicolumn{1}{|r|}{${\small 4}$} &
\multicolumn{1}{|r|}{${\small 11.6}$} & \multicolumn{1}{|r|}{${\small 12.4}$}
& \multicolumn{1}{|r|}{${\small 0.73}$} \\ \hline
\multicolumn{1}{|r|}{${\small Europe}$} & \multicolumn{1}{|r|}{${\small 16}$}
& \multicolumn{1}{|r|}{${\small 19.4}$} & \multicolumn{1}{|r|}{${\small 27.4}
$} & \multicolumn{1}{|r|}{${\small 0.73}$} \\ \hline
\multicolumn{1}{|r|}{${\small China}$} & \multicolumn{1}{|r|}{${\small 16}$}
& \multicolumn{1}{|r|}{${\small 25.5}$} & \multicolumn{1}{|r|}{${\small 32.1}
$} & \multicolumn{1}{|r|}{${\small 0.96}$} \\ \hline
\multicolumn{1}{|r|}{${\small Other}$} & \multicolumn{1}{|r|}{${\small 6}$}
& \multicolumn{1}{|r|}{${\small 14.5}$} & \multicolumn{1}{|r|}{${\small 17.5}
$} & \multicolumn{1}{|r|}{${\small 0.68}$} \\ \hline
\end{tabular}

{\small Table 2:} {\small The} {\small DSN$_{14}$ topological coefficients
averaged by partition of main destination.}
\end{center}

To have an idea on the influence of regional concerns in the patterns that
come out in the DSN$_{14}$, Figure 4 shows the same network of Figure 3 but
now the colors of the nodes are defined by their regional organizations:
blue for SADC countries, green for UMA, yellow for countries in the CEEAC,
red for those in COMESA and purple for the countries in CEDEAO.

At a first glance, the degree (number of links) of each country and the
general pattern of their connections do not seem to be conditioned by any
regional concern since the bulk of countries in the most connected part of
the network comprises countries of different regional organizations.
However, there is a slightly negative difference regarding COMESA (purple)
and CEDEAO (red) countries. Together with UMA countries (green) they are the
fewest connected countries in the entire DSN$_{14}$.

\begin{center}
\begin{figure}[htb]
\psfig{figure=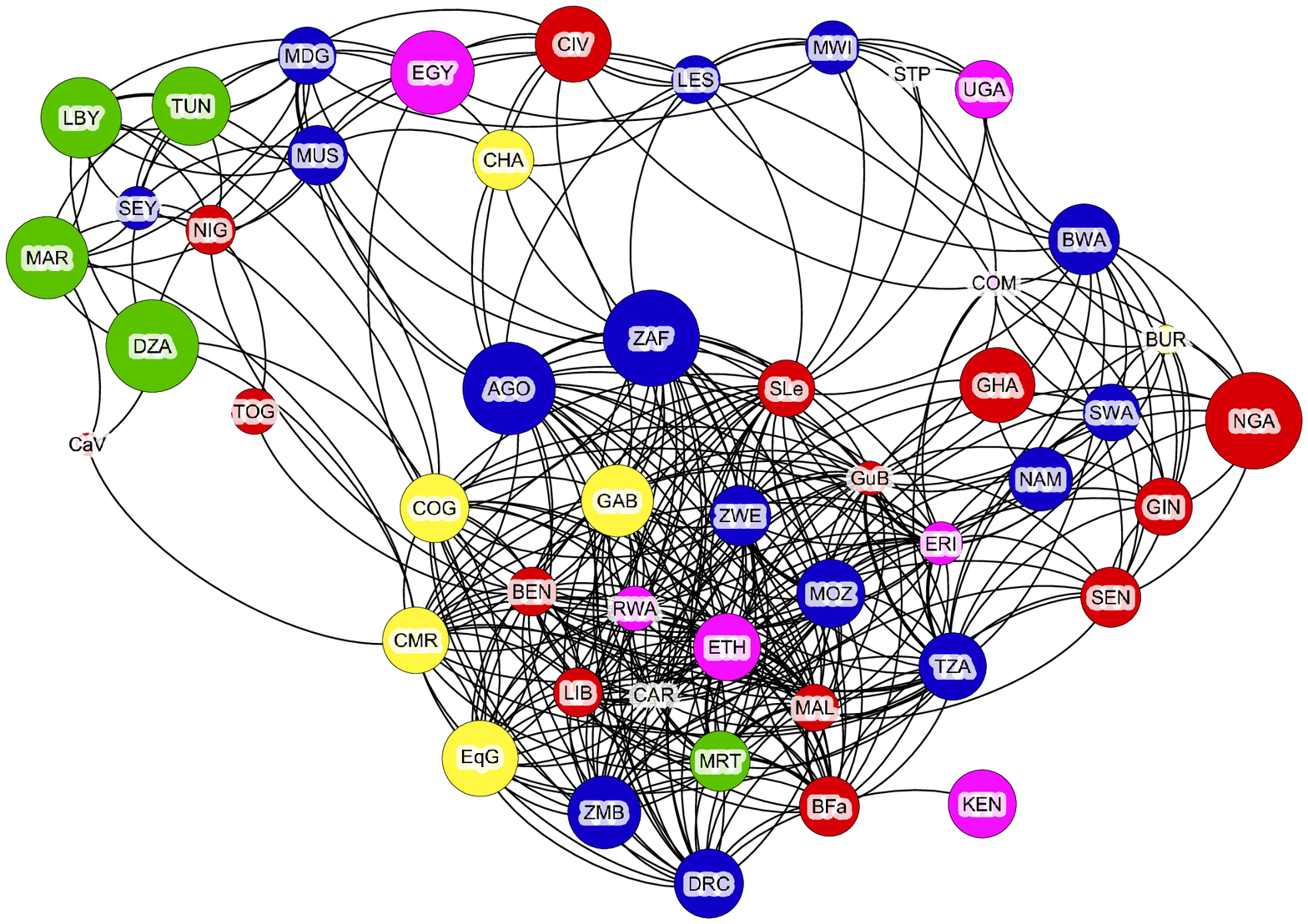,width=12truecm}
\caption{The DSN$_{14}$ colored by partition of regional
organization.}
\end{figure}
\end{center}

Table 3 shows some topological coefficients computed for each node of the DSN%
$_{14}$ and averaged by partition of regional organization. The averages of
the weighted degree ($\left\langle k\right\rangle $), betweenness ($%
\left\langle B\right\rangle $) and clustering ($\left\langle C\right\rangle $%
) of each partition show that although CEEAC partition (yellow) has just 8
countries, these countries have the greatest centrality ($\left\langle
k\right\rangle $ and $\left\langle B\right\rangle $) and clustering in the
whole DSN$_{14}$. This is certainly due to the fact that half of CEEAC
countries has "China"as their main export
destination and that besides exporting to China, the exports of CEEAC
countries are concentrated in a small number of other destinations.

Likely CEEAC countries, SADC members also display a high value of
betweenness centrality ($\left\langle B\right\rangle $), which is certainly
related to the fact that half of the countries in these regional
organizations occupy positions in the bulk of highly connected countries in
the "China"partition at the bottom of Figures
3 and 4. In contrast, UMA countries display low betweenness centrality,
showing that besides having "Europe"as their
main export destination, the second destination of exports of UMA countries
is spread over several countries.

\begin{center}
\begin{tabular}{|l|l|l|l|l|}
\hline
{\small Partition} & {\small Size} & $\left\langle k\right\rangle $ & $%
\left\langle B\right\rangle $ & $\left\langle C\right\rangle $ \\ \hline
\multicolumn{1}{|r|}{${\small SADC}$} & \multicolumn{1}{|r|}{${\small 15}$}
& \multicolumn{1}{|r|}{${\small 17.2}$} & \multicolumn{1}{|r|}{${\small 26.7}
$} & \multicolumn{1}{|r|}{${\small 0.7}$} \\ \hline
\multicolumn{1}{|r|}{${\small UMA}$} & \multicolumn{1}{|r|}{${\small 5}$} &
\multicolumn{1}{|r|}{${\small 12.2}$} & \multicolumn{1}{|r|}{${\small 6.5}$}
& \multicolumn{1}{|r|}{${\small 0.65}$} \\ \hline
\multicolumn{1}{|r|}{${\small CEEAC}$} & \multicolumn{1}{|r|}{${\small 8}$}
& \multicolumn{1}{|r|}{${\small 17}$} & \multicolumn{1}{|r|}{${\small 25.9}$}
& \multicolumn{1}{|r|}{${\small 0.86}$} \\ \hline
\multicolumn{1}{|r|}{${\small COMESA}$} & \multicolumn{1}{|r|}{${\small 7}$}
& \multicolumn{1}{|r|}{${\small 14.5}$} & \multicolumn{1}{|r|}{${\small 13}$}
& \multicolumn{1}{|r|}{${\small 0.62}$} \\ \hline
\multicolumn{1}{|r|}{${\small CEDEAO}$} & \multicolumn{1}{|r|}{${\small 14}$}
& \multicolumn{1}{|r|}{${\small 14.8}$} & \multicolumn{1}{|r|}{${\small 16.6}
$} & \multicolumn{1}{|r|}{${\small 0.83}$} \\ \hline
\end{tabular}

{\small Table 3: The DSN$_{14}$ topological coefficients averaged by
partition of regional organization.}
\end{center}

\subsubsection{Connecting countries by a mutual export commodity}

Here we develop the commodity share network (CSN$_{14}$) where African
countries in Table 1 are the network nodes and the intensity of a link
between each pair of them depends on the number of commodities that they
share as export products in 2014.

The bipartite network CSN$_{14}$ consists of the following partitions:

\begin{itemize}
\item the set $A$ of 49 African countries presented in Table 1 and

\item The set of commodities Commodities$_{14}$ (Section 2.2) that at least
one of the countries in the first partition have exported in 2014 on a first
and second commodity basis.
\end{itemize}

Therefore, in the CSN$_{14}$ two countries are linked if and only if they
shared a mutual leading export commodity in 2014. We have considered the two
main export products of each country in Table 1 (columns "Products"). Otherwise, if just the main product was
considered, the resulting CSN$_{14}$ would comprise a set of disconnected
sub-graphs as each country has just one main exporting product. Links in the
CSN$_{14}$ are weighted by the number of coincident products a pair of
countries share (among the two main products), consequently, every link $%
L_{(i,j)}$ in CSN$_{14}$ takes value in the set $\left\{ 0,1,2\right\} $.

As an example, the intensity of the link between KEN and UGA equals two ($%
L_{(KEN,UGA)}=2$) since KEN and UGA share two mutual leading export products
in 2014, they are Petroleum and Coffee. As another example, $L_{(ERI,RWA)}=1$
due to ERI and RWA mutual leading exports of Ores in 2014. Among the many
examples of missing links there are the cases of MOZ and KEN ($%
L_{(MOZ,KEN)}=0$) since MOZ and KEN did not share any mutual leading export
product in 2014.

Figure 5 presents the CSN$_{14}$, a network of 49 African countries linked
by at least one mutual leading export commodity in 2014. Nodes are colored
accordingly to the partition cluster to which their main exporting product
belongs: blue nodes have "Petroleum" as the
main export commodity in 2014, red nodes identify countries whose main
export products are "Manufactured", yellow for
"Diamonds", green for "Raw Materials" and purple for the cluster of "Other
Raw Materials".

Similarly to what was done within the DSN$_{14}$, the partition of the set
of countries into five clusters by main export commodity allows for
computing the average values of some topological coefficients by partition
cluster. In so doing it is possible to compare important patterns coming out
from the CSN$_{14}$. Like in the DSN$_{14}$ representations, the size of
each node is proportional to the export value of the country in 2014.

Likewise observed in the DSN$_{14}$, the graph in Figure 5 suggests the
existence of a positive and strong correlation between the amount of exports
of a country and its weighted degree in the CSN$_{14}$: the countries with
the highest amounts of exports in 2014 (the larger nodes) tend to be those
that cluster as petroleum exporters (blue), being followed by those that
export "Diamonds" (yellow) and "Manufactured"(red).

The highest degrees belong to AGO, NGA and ZAF, showing that these countries
are those that share (with other countries in the whole network) the highest
numbers of mutual export commodities. On the other hand, there are countries
like MWI and SWA that share just one exporting commodity with the all other
countries in the network.

In the CSN$_{14}$\ of Figure 5, the bulk of highly connected countries are
placed at the right side. It corresponds to the cluster of "Petroleum" exporters (blue), a highly clustered and almost
fully connected set of nodes. It is followed by the cluster of
"Diamonds" exporters (yellow) at the left side.
The very first pattern coming out from our CSN$_{14}$ is the remarkable
centrality of AGO.

\begin{center}
\begin{figure}[htb]
\psfig{figure=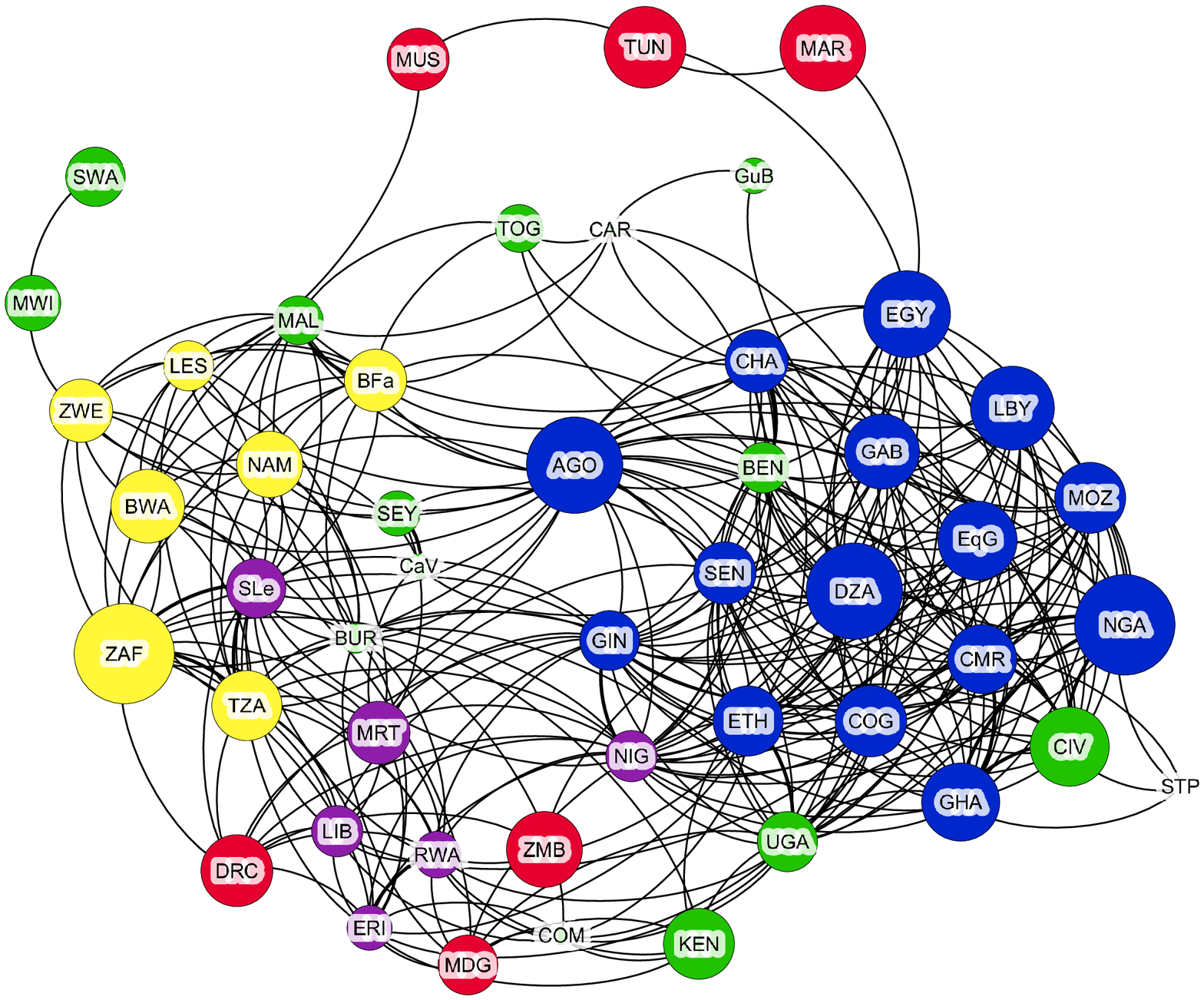,width=12truecm}
\caption{The CSN$_{14}$ colored by partition of main export
commodity.}
\end{figure}
\end{center}

Table 4 shows some topological coefficients computed for each node of the CSN%
$_{14}$ and averaged by partition of main exporting product. The averages of
the weighted degree ($\left\langle k\right\rangle $), betweenness ($%
\left\langle B\right\rangle $) and clustering ($\left\langle C\right\rangle $%
) of each partition show that the strongest connected countries are those
whose main exporting product is "Petroleum". In
terms of connectivity, they are followed by the cluster of Manufactured
Products, even though this cluster has just six countries.

There is a very high value of betweenness centrality ($\left\langle
B\right\rangle $) characterizing countries in the (small) "Diamonds" cluster, being certainly led by the centrality of
ZAF. Although this cluster has just seven countries, it displays the second
highest betweenness in the ranking of partitions. In so doing, we are
informed that the countries whose main exporting commodity is
"Diamonds" have a large sharing of common
export products with other countries in the whole network. The same applies
to the cluster of "Petroleum" exporters.

\begin{center}
\begin{tabular}{|l|l|l|l|l|}
\hline
{\small Partition} & {\small Size} & $\left\langle k\right\rangle $ & $%
\left\langle B\right\rangle $ & $\left\langle C\right\rangle $ \\ \hline
\multicolumn{1}{|r|}{${\small Petroleum}$} & \multicolumn{1}{|r|}{${\small 15%
}$} & \multicolumn{1}{|r|}{${\small 21.6}$} & \multicolumn{1}{|r|}{${\small %
35.5}$} & \multicolumn{1}{|r|}{${\small 0.82}$} \\ \hline
\multicolumn{1}{|r|}{${\small RawMaterials}$} & \multicolumn{1}{|r|}{$%
{\small 15}$} & \multicolumn{1}{|r|}{${\small 9.9}$} & \multicolumn{1}{|r|}{$%
{\small 14.1}$} & \multicolumn{1}{|r|}{${\small 0.72}$} \\ \hline
\multicolumn{1}{|r|}{${\small Diamonds}$} & \multicolumn{1}{|r|}{${\small 7}$%
} & \multicolumn{1}{|r|}{${\small 14.2}$} & \multicolumn{1}{|r|}{${\small %
33.7}$} & \multicolumn{1}{|r|}{${\small 0.70}$} \\ \hline
\multicolumn{1}{|r|}{${\small Manufactured}$} & \multicolumn{1}{|r|}{$%
{\small 6}$} & \multicolumn{1}{|r|}{${\small 18.6}$} & \multicolumn{1}{|r|}{$%
{\small 22.3}$} & \multicolumn{1}{|r|}{${\small 0.77}$} \\ \hline
\multicolumn{1}{|r|}{${\small OtherRM}$} & \multicolumn{1}{|r|}{${\small 6}$}
& \multicolumn{1}{|r|}{${\small 4.5}$} & \multicolumn{1}{|r|}{${\small 5.8}$}
& \multicolumn{1}{|r|}{${\small 0.48}$} \\ \hline
\end{tabular}

{\small Table 4:} {\small The CSN$_{14}$ topological coefficients averaged
by partition of main exporting product.}
\end{center}

Another interesting characteristic is the poor connectivity pattern of
countries in the "Raw Materials" partition
cluster. Even being a large partition in size, its average weighed degree ($%
\left\langle k\right\rangle $) is the second smallest in the CSN$_{14}$. It
means that countries that mainly export raw materials have a small share of
mutual export products with other countries.

\begin{center}
\begin{figure}[htb]
\psfig{figure=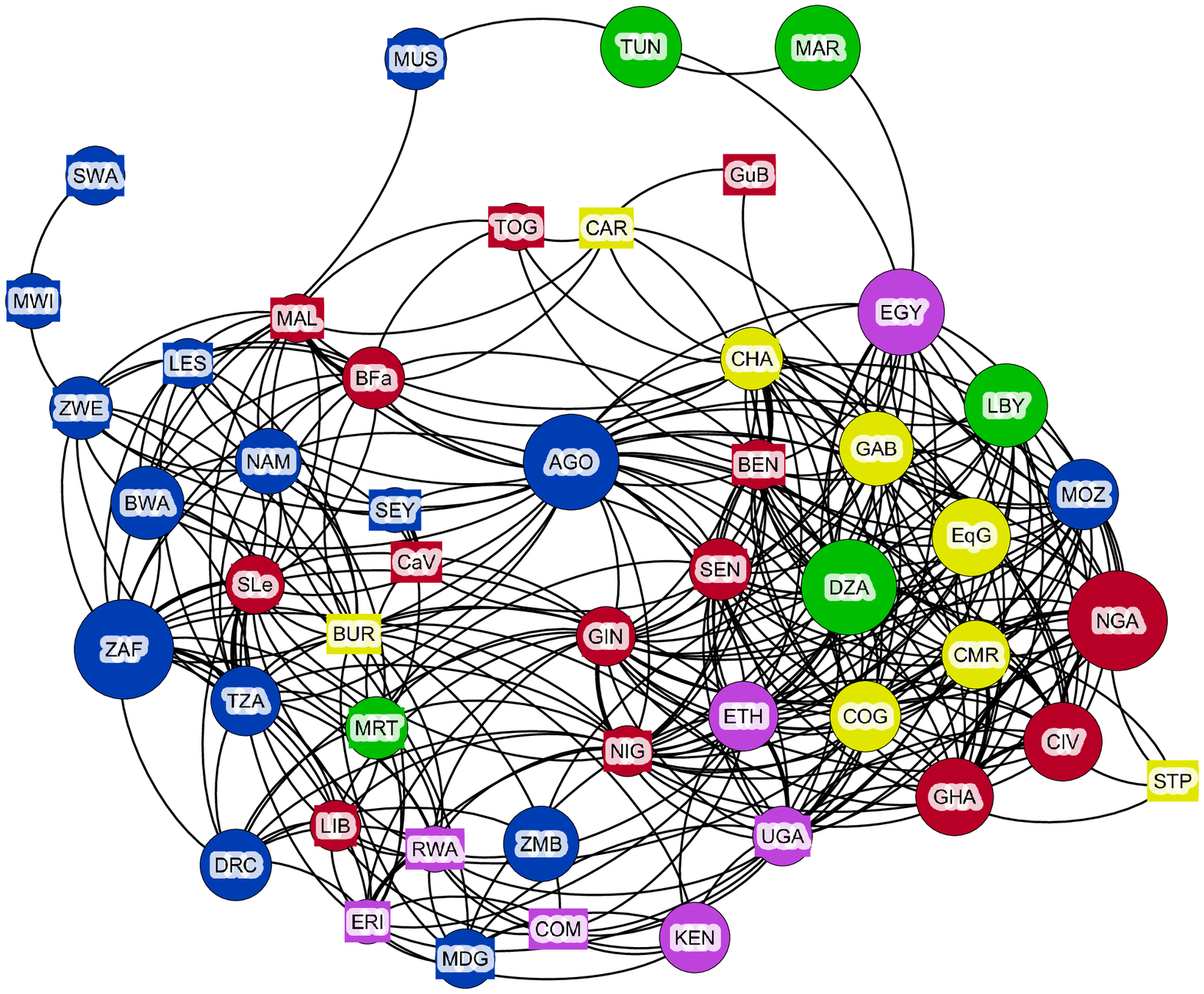,width=12truecm}
\caption{The CSN$_{14}$ colored by partition of regional
organization.}
\end{figure}
\end{center}

When a regional perspective is taken, the CSN$_{14}$ in Figure 6 is the same
network presented in Figure 5 but nodes are colored \ according to their
regional organizations: blue for SADC countries, green for UMA, yellow for
countries in the CEEAC, red for those in COMESA and purple for the countries
in CEDEAO.

Table 5 shows some topological coefficients computed for each node of the CSN%
$_{14}$ and averaged by partition of regional organization. There, the CEEAC
partition (yellow) although having just eight countries, has the greatest
clustering ($\left\langle C\right\rangle $) in the whole CSN$_{14}$. On the
contrary, the SADC cluster, although comprising a large number of countries,
is the one of the poorest clustering value in the ranking of five partitions
presented in Table 5.

Another evidence coming out from the regional perspective is that without
AGO and MOZ, the large set of fifteen SADC countries (blue) would be
excluded from the large cluster of "Petroleum" exporters. This is certainly related to the small clustering that SADC
countries have in the CSN$_{14}$, being most of SADC countries mainly
diamond and ore exporters.

\begin{center}
\medskip
\begin{tabular}{|l|l|l|l|l|}
\hline
{\small Partition} & {\small Size} & $\left\langle k\right\rangle $ & $%
\left\langle B\right\rangle $ & $\left\langle C\right\rangle $ \\ \hline
\multicolumn{1}{|r|}{${\small SADC}$} & \multicolumn{1}{|r|}{${\small 15}$}
& \multicolumn{1}{|r|}{${\small 10.7}$} & \multicolumn{1}{|r|}{${\small 27.7}
$} & \multicolumn{1}{|r|}{${\small 0.56}$} \\ \hline
\multicolumn{1}{|r|}{${\small UMA}$} & \multicolumn{1}{|r|}{${\small 5}$} &
\multicolumn{1}{|r|}{${\small 11}$} & \multicolumn{1}{|r|}{${\small 11.7}$}
& \multicolumn{1}{|r|}{${\small 0.76}$} \\ \hline
\multicolumn{1}{|r|}{${\small CEEAC}$} & \multicolumn{1}{|r|}{${\small 8}$}
& \multicolumn{1}{|r|}{${\small 16.3}$} & \multicolumn{1}{|r|}{${\small 19}$}
& \multicolumn{1}{|r|}{${\small 0.8}$} \\ \hline
\multicolumn{1}{|r|}{${\small COMESA}$} & \multicolumn{1}{|r|}{${\small 7}$}
& \multicolumn{1}{|r|}{${\small 16.8}$} & \multicolumn{1}{|r|}{${\small 29.6}
$} & \multicolumn{1}{|r|}{${\small 0.72}$} \\ \hline
\multicolumn{1}{|r|}{${\small CEDEAO}$} & \multicolumn{1}{|r|}{${\small 14}$}
& \multicolumn{1}{|r|}{${\small 16.9}$} & \multicolumn{1}{|r|}{${\small 21.4}
$} & \multicolumn{1}{|r|}{${\small 0.78}$} \\ \hline
\end{tabular}

{\small Table 5: The CSN$_{14}$ topological coefficients averaged by
partition of regional organization.}
\end{center}

\subsubsection{\textbf{Comparing the destination share and the commodity
share networks}}

Table 6 shows some network coefficients computed for the DSN$_{14}$ and the
CSN$_{14}$. The coefficients $\left\langle k\right\rangle $, $\left\langle
B\right\rangle $ and $\left\langle C\right\rangle $ were computed at the
node level and averaged by network. The averages of the weighted degree ($%
\left\langle k\right\rangle $), betweenness ($\left\langle B\right\rangle $)
and clustering ($\left\langle C\right\rangle $) of these networks show that
for the 49 African countries in 2014, sharing a mutual leading export
product happens less often than sharing a mutual leading destination of
exports, since the degree of the DSN$_{14}$ is greater than the degree of
the CSN$_{14}$.

The columns ("\textbf{density}") and (%
"\textbf{diameter}") provide values for the
most typical coefficients in network analysis. These coefficients are not
computed at the node level but for each (entire) network. Since the diameter
of the CSN$_{14}$ is larger than the diameter of the DSN$_{14}$, the 49
African countries are on average closer to each other when connected by a
mutual leading export destination than when connected by a mutual exporting
product.

\begin{center}
\begin{tabular}{|l|l|l|l|l|l|l|}
\hline
Graph & {\small Size} & $\left\langle k\right\rangle $ & $\left\langle
B\right\rangle $ & $\left\langle C\right\rangle $ & $density$ & $diameter$
\\ \hline
DSN$_{14}$ & ${\small 49}$ & \multicolumn{1}{|r|}{${\small 15.6}$} &
\multicolumn{1}{|r|}{${\small 19.7}$} & \multicolumn{1}{|r|}{${\small 0.76}$}
& \multicolumn{1}{|r|}{${\small 0.31}$} & \multicolumn{1}{|r|}{${\small 3}$}
\\ \hline
CSN$_{14}$ & ${\small 49}$ & \multicolumn{1}{|r|}{${\small 14.3}$} &
\multicolumn{1}{|r|}{${\small 23.1}$} & \multicolumn{1}{|r|}{${\small 0.72}$}
& \multicolumn{1}{|r|}{${\small 0.28}$} & \multicolumn{1}{|r|}{${\small 5}$}
\\ \hline
\end{tabular}

{\small Table 6: Comparing topological coefficients obtained for DSN$_{14}$
and CSN$_{14}$.}
\end{center}

The DSN$_{14}$ also has a larger clustering coefficient than the CSN$_{14}$,
showing that when a country shares a mutual export destination with other
two countries, these two other countries also tend to share a mutual export
destination between them. The densities of DSN$_{14}$ and CSN$_{14}$ confirm
that topological distances in the DSN$_{14}$ are shorter than in the CSN$%
_{14}$ and that on average going from one country in the DSN$_{14}$ to any
other country in the same graph takes less intermediate nodes than in the CSN%
$_{14}$.

Although the networks DSN$_{14}$ and CSN$_{14}$ inform about the degree of
the nodes, their densely-connected nature does not help to discover any
dominant topological pattern besides the distribution of the node's degree.
Moving away from a dense to a sparse representation of a network, one shall
ensure that the degree of sparseness is determined endogenously, instead of
by an a priory specification. It has been often accomplished (\cite%
{Araujo2012},\cite{Dias}) through the construction of a Minimal Spanning
Tree (MST), in so doing one is able to develop the corresponding
representation of the network where sparseness replaces denseness in a
suitable way.

\subsection{The Minimum Spanning Tree Approach}

In the construction of a MST by the \textit{nearest neighbor} method, one
defines the 49 countries (in Table 1) as the nodes ($N_{i}$) of a weighted
network where the distance $d_{ij}$ between each pair of countries $i$ and $%
j $ corresponds to the inverse of weight of the link ($d_{ij}=\frac{1}{L_{ij}%
}$) between $i$ and $j$.

From the $NxN$ distance matrix $D$, a hierarchical clustering is then
performed using the \textit{nearest neighbor} method. Initially $N$ clusters
corresponding to the $N$ countries are considered. Then, at each step, two
clusters $c_{i}$ and $c_{j}$ are clumped into a single cluster if

\begin{center}
$d\{c_{i},c_{j}\}=\min \{d\{c_{i},c_{j}\}\}$
\end{center}

with the distance between clusters being defined by

\begin{center}
$d\{c_{i},c_{j}\}=\min \{d_{pq}\}$ with $p\in c_{i}$ and $q\in c_{j}$
\end{center}

This process is continued until there is a single cluster. This clustering
process is also known as the \textit{single link method}, being the method
by which one obtains the minimal spanning tree (MST) of a graph.

In a connected graph, the MST is a tree of $N-1$ edges that minimizes the
sum of the edge distances. In a network with $N$ nodes, the hierarchical
clustering process takes $N-1$ steps to be completed, and uses, at each
step, a particular distance $d_{i,j}$ $\in $ $D$ to clump two clusters into
a single one.

\section{\protect\bigskip Results}

In this section we discuss the results obtained from the MST of each
one-mode projected graphs DSN$_{14}$ and CSN$_{14}$. As earlier mentioned,
the MST of a graph may allow for discovering relevant topological patterns
that are not easily observed in the dense original networks. As in the last
section, we begin with the analysis of the DSN$_{14}$ and then proceed to
the CSN$_{14}$.

We look for eventual topological structures coming out from empirical data
of African exports, in order to see whether some relevant characteristics of
African trade have any bearing on the network structures that emerge from
the application of our approach. In the last section, we observed some
slight influence of the regional position of each country in its
connectivity. With the construction of the minimum spanning trees we
envision that some stronger structural patterns would come to be observed on
the trees.

\subsection{The MST of the destination share network}

Figure 7 shows the MST obtained from the DSN$_{14}$ and colored according to
each country main destination of exports in 2014.

The first evidence coming out from the MST in Figure 7 is the central
position of AGO clustering together the entire set of "China" exporters (red) in 2014. Another important pattern that
emerges in the MST is the branch of UMA countries (yellow) in the right side
of the tree, being "Europe" their most frequent
destination of exports. Similarly, part of the countries that export mostly
to "Other" seems to cluster on the left branch
(purple). Interestingly, the countries that exports to "African countries" (blue) occupy the less central positions on
the tree. This result illustrates the suitability of the MST to separate
groups of African countries according to their main export destinations and
the show how opposite are the situations of those that export to
"China" from the countries that have Africa
itself as their main export destinations.

\begin{center}
\begin{figure}[htb]
\psfig{figure=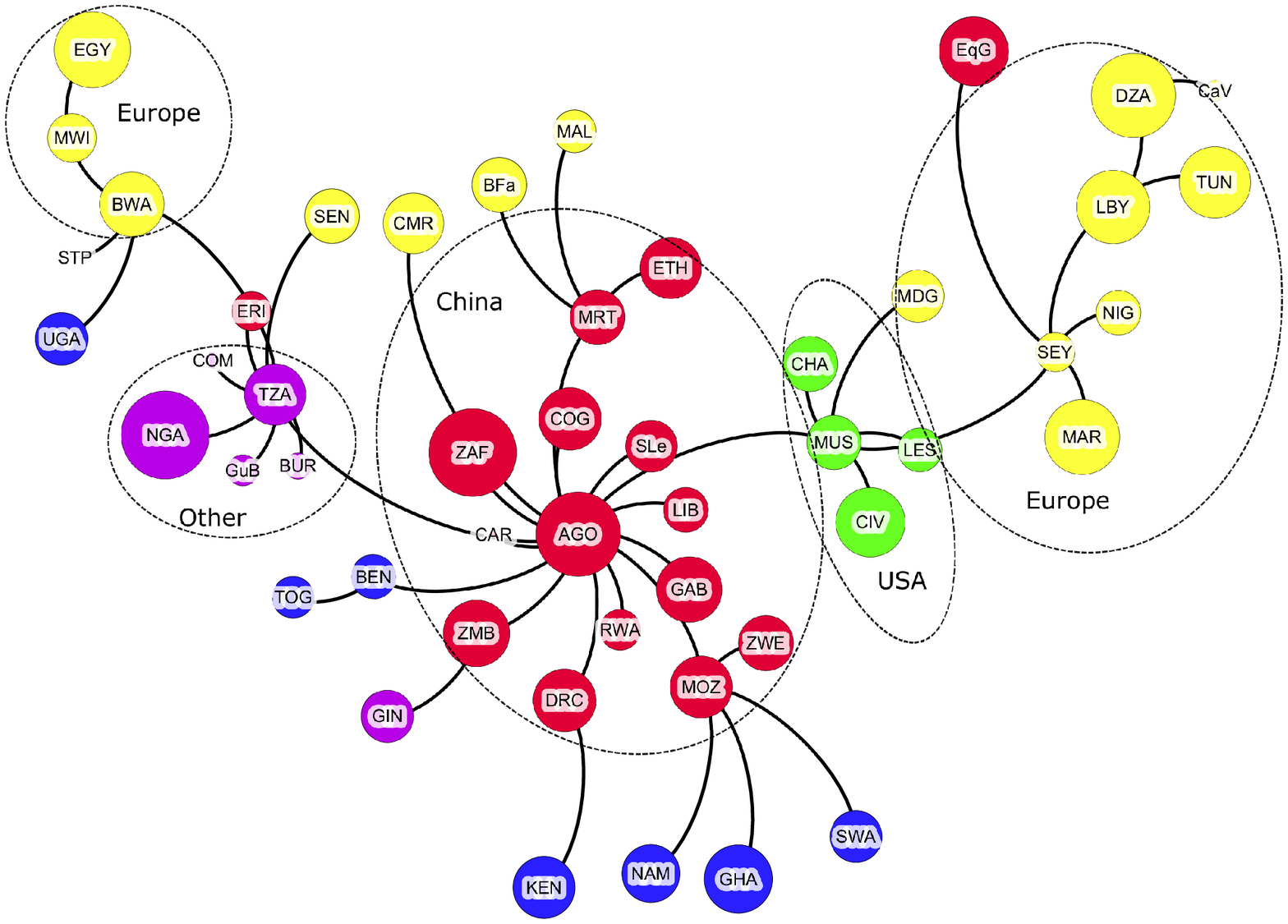,width=12truecm}
\caption{The MST of the CSN$_{14}$ colored by partition of main
destination.}
\end{figure}
\end{center}

Regarding centrality, AGO occupies the most central position of the network
since this country exports to the top most African export destinations (%
"China" and "Europe") being therefore, and by this means, easily connected to a
large amount of other countries. Indeed, AGO is the center of the most
central cluster of "China" exporters. On the
other hand, many leaf positions are occupied by countries that exports to
other African countries as they have the smallest centrality in the whole
network, they are KEN, NAM and SWA. Their weak centrality is due to the fact
that their leading export destinations are spread over several countries
(ZMB, TZA, ZAF, BWA and IND).

\subsection{The MST of the commodity share network}

Figure 8 shows the MST obtained from the CSN$_{14}$ and colored according to
the main export commodity of each country in 2014. The first observation on
the MST presented in Figure 8 is that, centrality is concentrated in a fewer
number of countries (when compared to the MST of the DSN$_{14}$).

\begin{center}
\begin{figure}[htb]
\psfig{figure=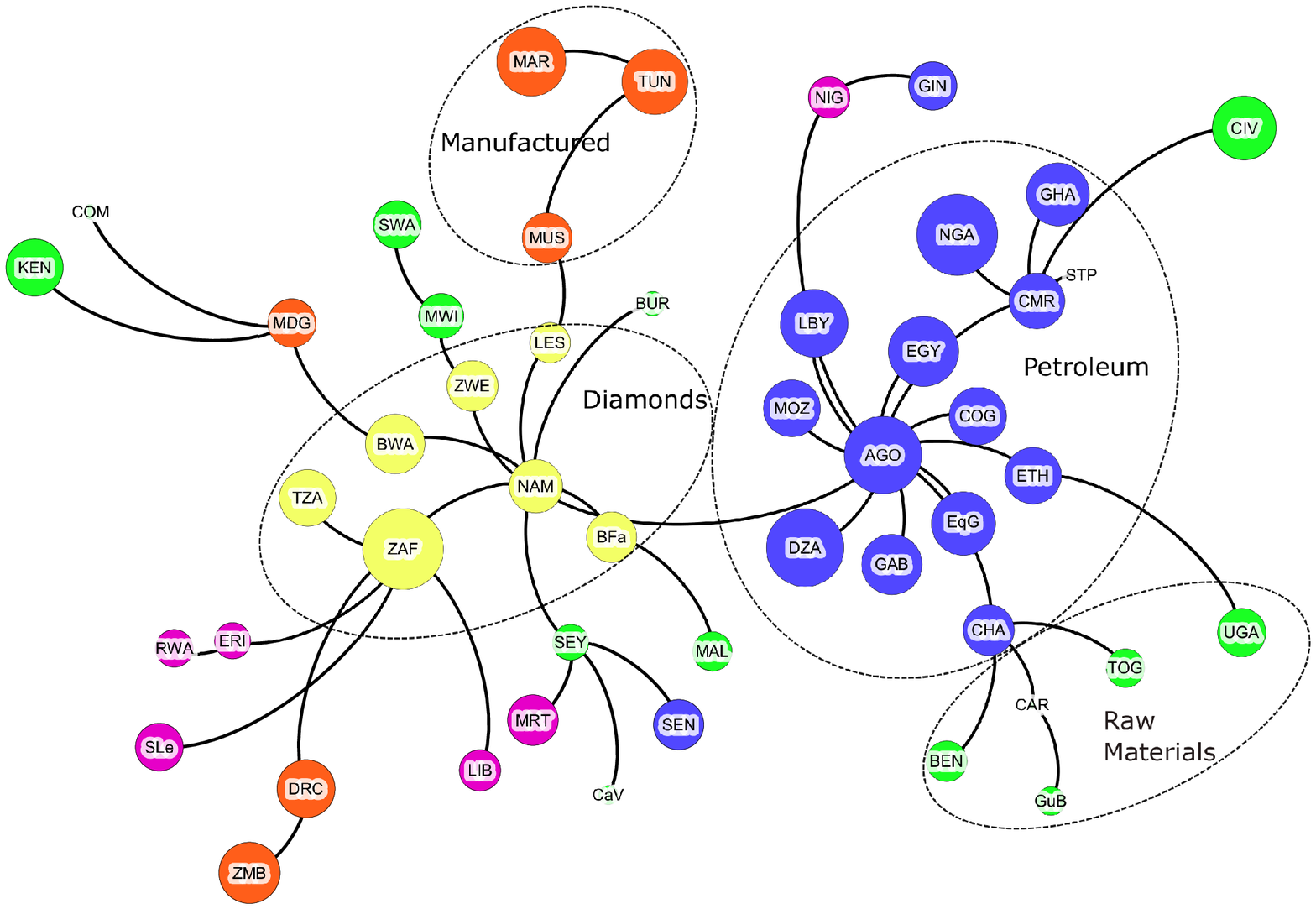,width=12truecm}
\caption{The MST of the CSN$_{14}$ colored by partition of main exporting commodity.}
\end{figure}
\end{center}

The top most central and connected positions are shared by countries
belonging to two regional organizations: SADC and CEDEAO, being mainly
represented by ZAF and AGO and clustering countries whose main export
commodities are "Diamonds" and "Petroleum", respectively. Unsurprisingly, centrality and
connectivity advantages seem to be concentrated in these two leading
commodity partitions ("Diamonds" and
"Petroleum") and organization groups (SADC and
CEDEAO).

Indeed, half of CEDEAO countries\ occupy the upper branch rooted in ZAF and
having "Diamonds"(yellow) as their main export
commodity. Another regional cluster is rooted in AGO and tie together
several UMA countries whose main export commodity is "%
Petroleum"(blue). On the other hand, half of UMA countries
are far from each other on the tree, they occupy the leaf positions, being
weakly connected to the other African countries to which, the few
connections they establish rely on having "Manufactured" as their main export commodity. Likewise, there is a branch
clustering exporters of "Raw Materials"(green)
being also placed at the leaf positions on the tree. Such a lack of
centrality of "Raw Materials" exporters in the
CSN$_{14}$ seems to be due to the fact that their leading export products
are spread over many different commodities (the "Raw Materials" partition comprises 12 different commodities).

\section{\protect\bigskip Concluding remarks}

In the last decade, a debate has taken place in the network literature about
the application of network approaches to model international trade. In this
context, and even though recent research suggests that African countries are
among those to which exports can be a vehicle for poverty reduction, these
countries have been insufficiently analyzed.

We have proposed the definition of trade networks where each bilateral
relation between two African countries is defined from the relations each of
these countries hold with another entity. Both networks were defined from
empirical data reported for 2014.They are independent bipartite networks: a
destination share network (DSN$_{14}$) and\ a\ commodity share network (CSN$%
_{14}$). In the former, two African countries are linked if they share a
mutual leading destination of exports, and in the latter, countries are
linked through the existence of a mutual leading export commodity between
them.

Our conclusions can be summarized in the following.

\begin{enumerate}
\item Sharing a mutual export destination happens more often: The very first
remark coming out from the observation of both the DSN$_{14}$ and\ the CSN$%
_{14}$ is that, in 2014 and for the 49 African countries, sharing a mutual
exporting product happens less often than sharing a mutual destination of
exports.

\item Great exporting countries tend to be more linked: There is a positive
correlation between strong connected countries in both the DSN$_{14}$ and
the CSN$_{14}$ and those with high amounts of export values in 2014. It is
in line with recent research placed in two different branches of the
literature on international trade: the World Trade Web (WTW)\ empirical
exploration (\cite{Serrano},\cite{Almog},\cite{Fagiolo},\cite{Saracco}, \cite%
{Benedictis},\cite{Picciolo},\cite{Yang}) and the one that specifically
focus on African trade (\cite{Portugal-Perez},\cite{Kamuganga},\cite%
{Morrissey},\cite{Baliamoune},\cite{Ackah},\cite{Gamberoni}). References (%
\cite{Kamuganga},\cite{Baliamoune}) reports on the role of export
performance to economic growth. They also discuss on the relation between
trade and development, and on the growth by destination hypothesis,
according to which, the destination of exports can play an important role in
determining the trade pattern of a country and its development path.

\item Destination matters: The idea that destination matters is in line with
our finding that in the DSN$_{14}$, the highest connected nodes are those
whose main export destination is China. According to Baliamoune-Lutz (\cite%
{Baliamoune}) export concentration enhances the growth effects of exporting
to China, implying that countries which export one major commodity to China
benefit more (in terms of growth) than do countries that have more
diversified exports.

\begin{itemize}
\item The China effect: One of the patterns that came out from our DSN$_{14}$
shows that half of CEEAC and SADC countries belongs to the bulk of
"China" destination cluster, having high
betweenness centrality. Additionally, the "China" destination group of countries displays the highest
clustering coefficient (0.96), meaning that, besides having China as their
main exports destination, the second destination of exports of the countries
in this group is highly concentrated on a few countries.

\item The role of Intra-African trade: Another important pattern coming out
from the MST of the DSN$_{14}$ shows how opposite are the situations of the
countries that export to "China" from the
countries that have Africa itself as their main export destinations. In the
MST of the destination share network, many leaf positions are occupied by
intra-African exporters as they have the smallest centrality in the whole
network. Their weak centrality is due to the fact that their leading export
destinations are spread over several countries. This result is in line with
the results reported by reference (\cite{Kamuganga}) where the growing
importance of intra-African trade is discussed and proven to be a crucial
channel for the expansion of African exports. Moreover, Kamuganga found
significant correlation between the participation in intra-African trade and
the diversification of exports.

\item The Angola cluster: our results highlighted the remarkable centrality
of AGO as the center of the most central cluster of "China" exporters. Indeed, AGO is the country that holds the most
central position when both DSN$_{14}$ and CSN$_{14}$ are considered. This
country occupies in both cases the center of the largest central clusters:
"China" exporters in DSN$_{14}$ and exporters
of "Petroleum" in CSN$_{14}$.

\item UMA countries anti-diversification: In the opposite situation, we
found that UMA countries display very low centrality, showing that besides
having "Europe" as their first export
destination, the second destination of exports of UMA countries is spread
over several countries. This result is in line with reference (\cite%
{Gamberoni}) report on European unilateral trade preferences and
anti-diversification effects. We showed that UMA countries occupy a separate
branch in the MST of the DSN$_{14}$, being "Europe" their most frequent destination of exports.
\end{itemize}

\item[4.] In the CSN$_{14}$, the highest connected nodes are those that
cluster as "Petroleum" exporters, being
followed by those that export "Diamonds".
Unsurprisingly, "Raw Materials" exporters
display very low connectivity as their second main exporting product is
spread over several different commodities.

\item[5.] Organizations matter: Regional and organizational concerns seem to
have some impact in the CSN$_{14}$.

\begin{itemize}
\item SADC and Petroleum: The group of SADC countries, although comprising a
large number of elements, is the one with the poorest connectivity and
clustering in the CSN$_{14}$. It is certainly due to the fact that without
AGO and MOZ, this large group of countries does not comprise "Petroleum" exporters.

\item UMA countries anti-diversification: Again in the CSN$_{14}$, its MST
shows that UMA countries are placed on a separate branch. Although they are
countries with high amounts of export values in 2014, UMA countries display
low connectivity and low centrality. The leaf positions in the MST of either
DSN$_{14}$ or CSN$_{14}$ - while occupied by countries with very low
centrality and connectivity - were shown to characterize countries that
export mainly to "Europe" and whose main
exporting product is "Raw Materials".
\end{itemize}
\end{enumerate}

Future work is planned to be twofold. We plan to further improve the
definition of networks of African countries, enlarging the set of
similarities that define the links between countries in order to include
aspects like mother language, currencies, demography and participation in
trade agreements. On the other hand, we also plan to apply our approach to
different time periods. As soon as we can relate the structural similarities

(or differences) and their evolution in time to certain trade
characteristics, the resulting knowledge shall open new and interesting
questions for future research on African trade.

\end{document}